\documentclass[oribibl]{llncs}

\usepackage[xdvi]{graphicx}
\usepackage{epsfig}

\begin{document}
\bibliographystyle{plain}
\frontmatter          

\title{Model of self-replicating cell capable of self-maintenance.}

\author{
Naoaki Ono
 and
Takashi Ikegami}
\institute{Institute of Physics, The Graduate School of Arts and
  Sciences,\\
University of Tokyo, 3-8-1 Komaba, Meguro-ku,Tokyo 153, Japan\\ 
\email{nono@sacral.c.u-tokyo.ac.jp}\\
\email{ikeg@sacral.c.u-tokyo.ac.jp}
}

\maketitle

\begin{abstract}
We have constructed a simple model of a proto-cell that simulates
stochastic dynamics of abstract chemicals on a two-dimensional lattice.
We have assumed that chemicals catalyze their reproduction through interaction
with each other, and that repulsion occurs between some chemicals.
We have shown that chemicals organize themselves into a cell-like
structure that maintains its membranes dynamically.
Further, we have obtained cells that can divide themselves
automatically into daughter cells. 
\end{abstract}

\section{Introduction}
The emergence of cells is one of the major transitions in the
evolution of life.
When primitive self-replicators such as a hypercycle of
RNA enzymes evolve into a living cell, they must acquire
membranes that will separate them from their noisy environment. 
It is well known that a hypercycle system can easily be
broken down by the occurrence of parasites.
Compartmentalization of a hypercycle system is a simplest way to 
avoid the disaster \cite{Eigen1981} \cite{Szathmary1987}
\cite{Szathmary1997}. 
At the same time, however, it should be noticed that it is also true that  
parasites can drive the increase of diversity and
complexity of the replicator network \cite{Ikegami1997}.
In order to examine the balance between stable reproduction and diversity, 
we should study a relationship
between an internal replication and a cellular structure which
enclose it. 

Many models of proto-cell structures have been proposed. 
For example, it is well known that long-chained fatty acids
spontaneously form micelles or vesicles when submerged in water.
Luisi and his group demonstrated experimentally the
self-organization and self-reproduction of liposomes, and showed
that such 
vesicles maintain self-replicating RNA within them \cite{Oberholzer1995}.
Theoretical models for self-organization and
self-reproduction of micelles are
studied well \cite{Coveney1996} \cite{Rasmussen1997}.

There is an another essential feature of cells, that is,
self-maintenance.
Living cells metabolize and sustain their membrane by themselves,
and the boundary of cells are defined by the membrane.
This mutual dependence enables the coevolution of internal chemical
networks and the membranes.
The coevolution between the two is presumed in the early
stages of the cell evolution. 
With respect to this point, 
G\'anti proposed a model for primitive life termed 'the
chemoton', which presents three indispensable functions of the
proto-cell: 
it has a metabolic cycle for assimilation; it
maintains its membrane; and it replicate its genetic
information \cite{Ganti1975} \cite{Ganti1997}.
Varela also insisted that the boundary of 
cells (i.e. the cell membrane) must be organized and maintained by
the cell itself \cite{Maturana1980} \cite{Maturana1987}.
He presented a model on a two-dimensional lattice of an
autopoietic cell that can maintain its membrane.

Our purpose in the present study is to demonstrate how such primitive 
cells can emerge and evolve from a simple set of
chemical network. 
A model of self-maintenance and self-replicating cells in
one-dimensional 
space was proposed by the same author \cite{Ono1999}.
In the model, we have shown that
self-reproduction of the cellular structure emerges 
spontaneously and there are two distinct processes of 
replication showing potentially different heredities.

In this paper, we extend our previous model 
for application to two-dimensional cases
and showing that this system has a 
potential for further evolution.

\section{A stochastic particle model}

We simulate a discrete space-time dynamics of chemicals in a
two-dimensional space, where chemicals catalyze each other's reactions.
Each chemical is given as a particle with/without anisotropic shape that
moves around on a triangular lattice. Particles demonstrate two basic
motions:  hopping to neighbouring sites and rotating at
one site.  In addition to this behavior, a particle can
change its chemical qualities. 
The former is termed a mobile transition, the latter a chemical
transition, and both are
determined by the potential energy of the particle.

We assume that there is a repulsive force between some chemicals,
thus the physical potential of a chemical $C$ at
the site  $x$ ($E_C(x)$) is computed by summing up the the repulsion 
potential of all chemicals at the site $x$ and its six
neighbouring sites. 
The mobile transition probability $P_C(x,x_0)$ from
site $x$ to $x_0$ is computed from the difference in the
potential magnitudes as 

$$ P_C(x,x_0) = R_{dif} \> f(E_C(x_0)-E_C(x)), $$

where $E_C(x)$ gives the potential energy of the particle $C$ at the
site $x$. The diffusion  parameter $R_{dif}$ is fixed for all particles.

The chemical transition probability $P_{C \rightarrow C'}(x)$ from 
the state $C$ to $C'$ at the site $x$ is given as

$$ P_{C \rightarrow C'}(x) = R_{C \rightarrow C'}(x) \> 
f(G_{C'}- G_C + E_{C'}(x)-E_C(x)), $$

where $G_C$ represents  the chemical potential of $C$. 
The reaction parameter $R_{C \rightarrow C'}(x)$ is controlled
by a catalyst found on the site $x$.
One constraint is given to the form of the function $f$ 
in order to satisfy the 
thermal equilibrium condition, as follows:

\begin{equation}
  \label{thermal}
  \frac{f(\Delta E)}{f(-\Delta E)} = e^{-\Delta E}
\end{equation}

We define five different kinds of chemicals, $A; M; W; X$ and $Y$,
in a system.
Each particle can belong to any one of these chemicals.  
$W$ plays the role of abstract 'water', and cannot change into
any other chemical. 
$X$ is the material with 
high chemical potential, though it is not an autocatalytic
chemical. 
$A$ is a unique autocatalytic chemical in the system. Their
reaction processes are,

\begin{eqnarray*}
 A + A  &\leftrightarrow& AA \\
  and\\
 X + AA &\leftrightarrow& A + AA.
\end{eqnarray*}

These chemical reactions only occur among particles that occupy
the same site
\footnote{Note that a number of chemicals can occupy each site. In this
study, the average is one hundred.}
The forward and backward reactions have an equal reaction
parameter, which is given by the following formula:

$$ R_{X \rightarrow A}(x) = R_{A \rightarrow X}(x) = B_{X \leftrightarrow A} +
C_{A} A(x)^2,$$

where $A(x)$ denotes the number of  $A$ on the site $x$.
In the above equation, $B_{X \leftrightarrow A}$ and $C_{A}$ are
the base rate and the 
catalysis coefficient, respectively.  
As a secondary process, $A$ produces $M$ as a co-product of the total
reaction network.

\begin{eqnarray*}
 X + AA &\leftrightarrow& M + AA.
\end{eqnarray*}

The reaction parameters are given by

$$ R_{X \rightarrow M}(x) = R_{M \rightarrow X}(x) = 
B_{X \leftrightarrow M} + C_{M} A(x)^2. $$ 

In addition to the above reactions, we introduce the natural decay
of chemicals into $Y$ where $Y$ has the lowest
chemical potential.
 
Consider that there is a source of material $X$ in this system
so that the reaction parameters between $X$ and $Y$ break 
the pattern of symmetry, as 

\begin{eqnarray*}
 R_{X \rightarrow Y} &=& B_{X \leftrightarrow Y} \\
 R_{Y \rightarrow X} &=& B_{X \leftrightarrow Y} + S_x 
\end{eqnarray*}

where $S_x$ denotes the strength of the source $X$.

We assume there is repulsive force between $M$ and other
chemicals like oil in water. 
In the following simulations, we examine three different kinds
of potentials on the chemical $M$.
First, $M$ equally repels all the other chemicals around it.
Second, $M$ has an anisotropic repulsion  regardless of the kinds of
chemicals.  This feature  will be described latter.
Third, in addition to the anisotropic repulsion, the repulsion
force also depends on the kinds of chemicals. 

\section{Simulation Results}

\subsection{Formation of Cells}

First, we simulate the case where the repulsion force depends
on neither the kinds of chemicals nor the form of $M$.
Starting from the homogeneous initial state which has rich
amount of $A$, a system can maintain replication of $A$ and
reproducing co-product $M$. 
Chemicals $A$ and $M$ can aggregate to form  Turing-like
patterns. 
Figure~1.a shows an example of the pattern generated; the spots
of $M$ are formed among $W$ and $A$.

The second case, where the repulsion force depends on the
orientation of $M$ molecule gives different observation.
Here $M$ has an anisotropic potential, illustrated in Fig.~1.b. 
When these $M$s are placed on  a triangular lattice, the 'head'
can be aligned in any of three directions (e.g. $0$,$\pi/3$ and
$2\pi/3$). We assume that it can change its direction
stochastically, with the transition probability given by,
$$ P_{MM'}(x) = R_{rot} \> f(E_M(x)-E_M'(x)), $$
where $M'$ denotes $M$ which has another orientation.

Figure~1.b shows the repulsion potential generated by $M$ for other
chemicals $A$,$W$,$X$ and $Y$.
The repulsion force from $M$ becomes strongest when  $M$ and
other chemicals are 
on the same site (indicated by black in Fig.~1.b). 
The repulsion is the second strongest at the front of or behind
M $M$ (dark gray sites) and relatively weaker at 
the other four side-sites (light gray).
We also assume there is repulsion between $M$s when their
directions are different. 
Thus, the $M$ molecule tends to take the same direction as
neighbouring $M$s.

When $M$ has this kind of anisotropic repulsive force,
the clusters are organized differently than they are in the
isotropic cases. 
The clusters of $M$ become thin films that we simply name
'membrane' (see Fig.~1.c). 
The difference in the repulsion potential between front- and
side-sites of $M$ 
affects the thickness of membrane. Also, when the repulsion
between $M$ with different direction is stronger, the
membranes tend to run straighter. These effects allow us to get
membranes that have various degrees of flexibility.

When we start from a single 'cell', 
that is, a spot of $A$ enveloped by membranes as shown Fig.~1.d,  
this structure can maintain itself stably 
because $A$ within the cell keep reproducing themselves
and sustain the membranes by supplying $M$; simultaneously, the
membranes keep $A$ from diffusing outward.
Note that this structure collapses when the membranes are
broken (see Fig.~1.e). Chemical $A$ cannot sustain
reproduction because they leak away through the defect in the
membrane. 
In the absence of a supply by $A$, the membranes decay and
disappear.

\begin{figure}[h]
  \begin{center}
    \begin{tabular}{ccc}
      \includegraphics[scale=0.9]{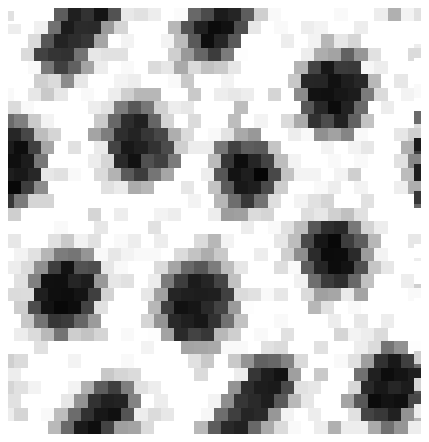} &
      \includegraphics[height=3.2cm]{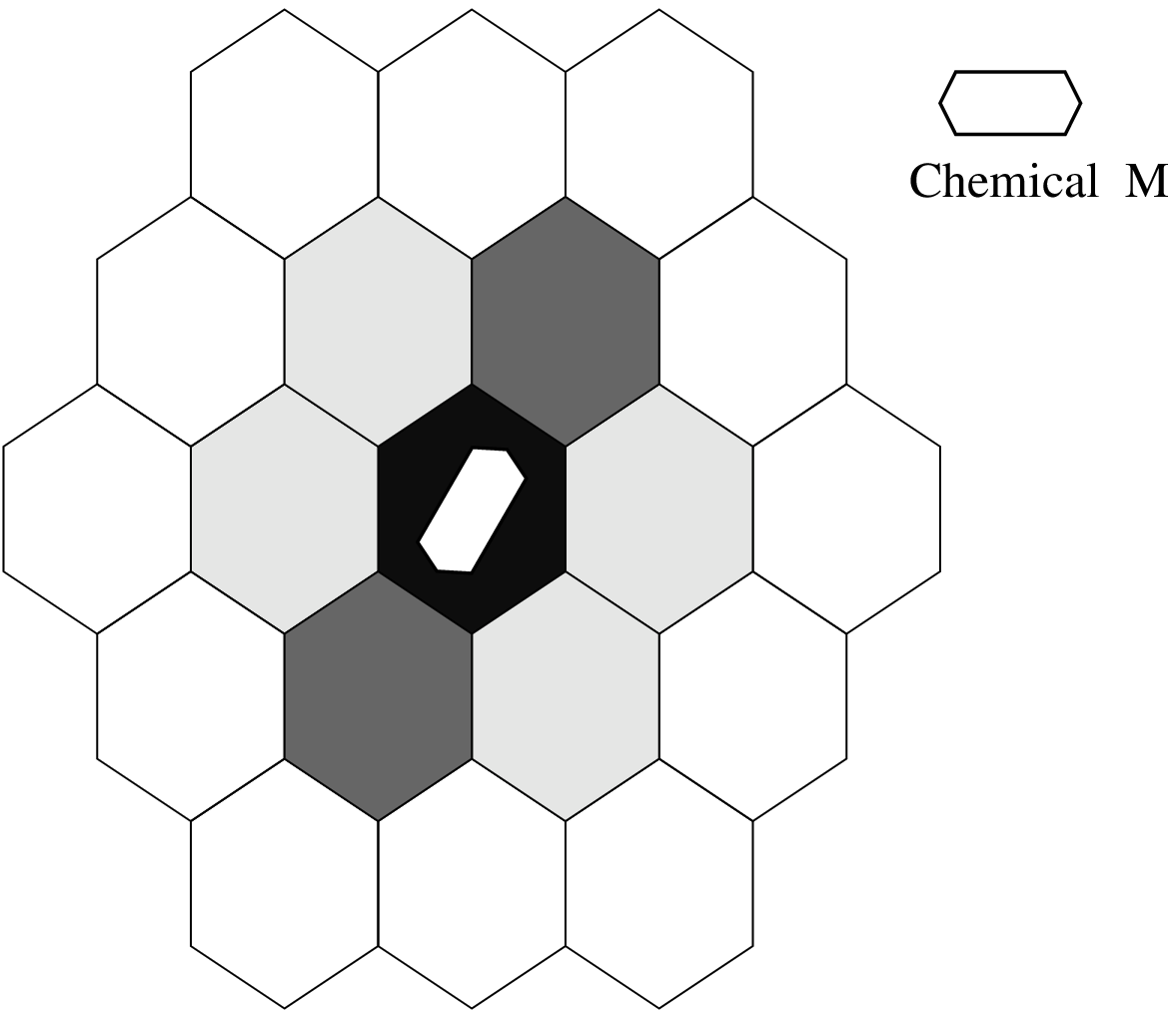} &
      \includegraphics[scale=0.9]{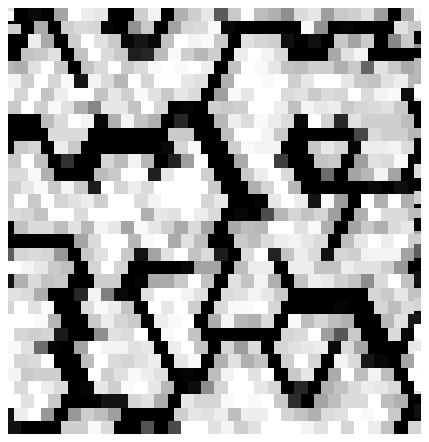} \\
      (a) & (b) & (c) \\
      \includegraphics[scale=0.9]{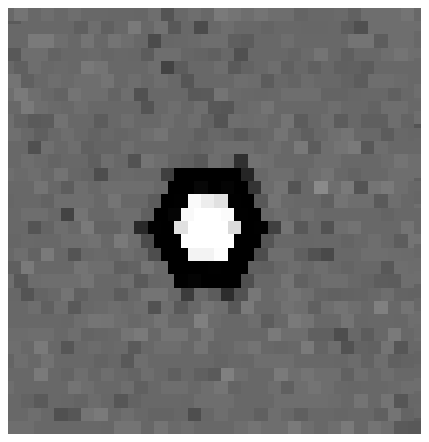} &
      \includegraphics[scale=0.9]{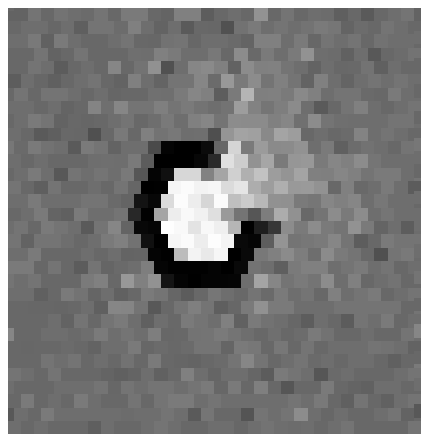} &
      \includegraphics[scale=0.9]{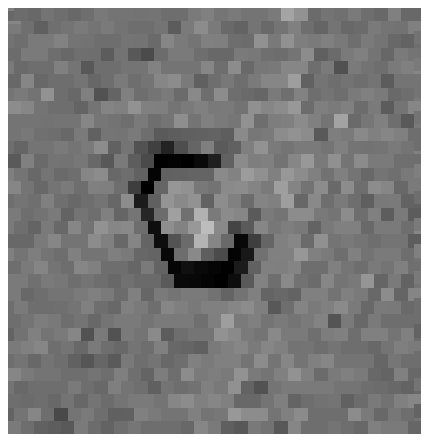} \\
      (d) & (e.1) t=3000 & (e.2) t=6000
    \end{tabular}    
  \end{center}
  \label{clusters}
\caption{
Cluster formation.
Picture (a) presents an example of the patterns formed by
molecules with isotropic repulsion.
Picture (b) illustrates the anisotropic field of repulsion
around $M$. The depth of gray denotes the intensity of
repulsion. 
Picture (c) shows that $M$ forms thin films and separates the
domains of $A$. 
The white domain contains rich amount of $A$. The gray and black
domains are dominated by $W$ and $M$, respectively. 
In picture (d), a cell structure maintains itself stably. 
Pictures 
(e.1) and (e.2) are the snapshots of the collapse of a cell
starting from a cell which lacks the upper right membrane.}
\end{figure}

\subsection{Cell division}

Living cells are not closed systems. They must ingest
nutrients and excrete wastes through their membranes.
In this section, we study the case where $M$ shows selective
repulsion depending on the kind of molecules.
We assume that the repulsion between $M$ and two chemicals $X$ and
$Y$ is much weaker than that of the other chemicals.

In this case, $X$ and $Y$ can permeate through the membranes at
a rate proportional to the gradient of their density.
Because there are more $X$ in the environment than on the inside
of the cell, the cells can absorb the external chemical $X$ and
grows gradually. 
When the cell reaches a certain size at which it has outgrown 
its stability, it begins to 
generate a new membrane inside. This finally divides
the mother cell into daughter cells (see Fig.~2).
These new cells repeat the process of growing and dividing.
Sometimes a cell fails to sustain its membrane structure and
dies, due to a shortage of materials or to interference from
other cells. 

We can change the flexibility of the membranes by
altering the repulsion strength  between $M$s. This has the
result of varying the division dynamics of cells. Examples are
presented in Fig.~3.
A strong repulsion between $M$ results in the formation of stiff
membranes, and the shape of cells  becomes more 
regular (Fig.~3.a and 3.b). 
On the other hand, Fig.~3.c represents a cell with more flexible
membranes, which form in the presence of low repulsion values for $M$.
These cells divide themselves irregularly at some narrow part.

\begin{figure}[htbp]
  \begin{center}
    \begin{tabular}{ccc}
      \includegraphics[scale=0.9]{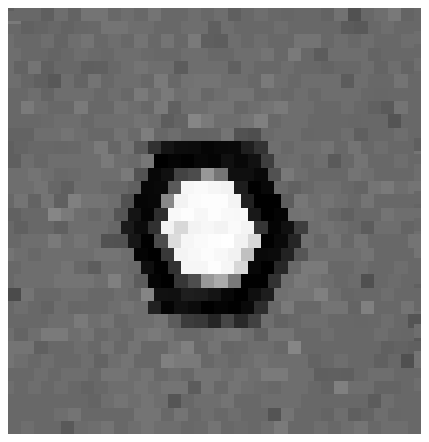} & 
      \includegraphics[scale=0.9]{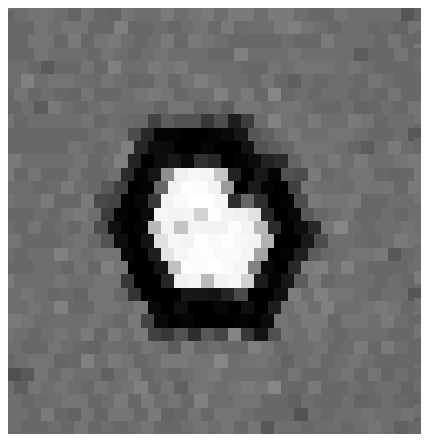} &
      \includegraphics[scale=0.9]{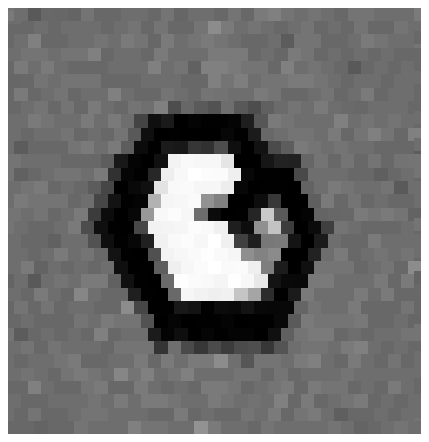} \\
      (a.1) t = 3000 & (a.2) t = 6000 & (a.3) t = 9000 \\
      \includegraphics[scale=0.9]{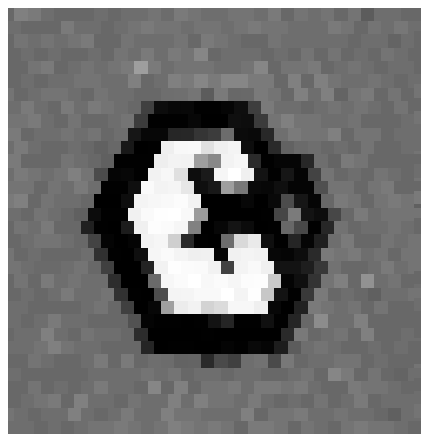} &
      \includegraphics[scale=0.9]{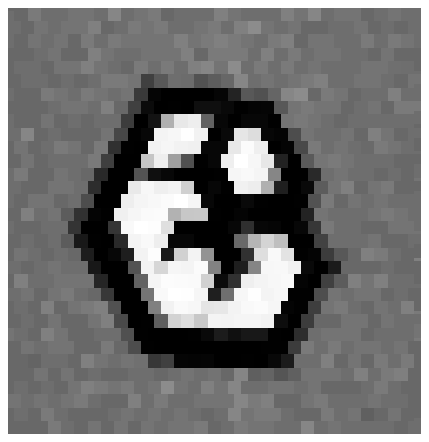} &
      \includegraphics[scale=0.9]{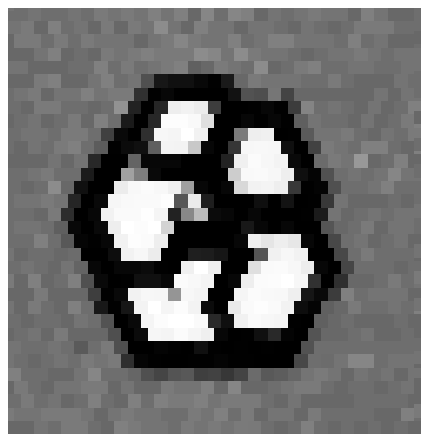} \\
      (a.4) t = 12000 & (a.5) t = 15000 & (a.6) t = 18000 
    \end{tabular}    
  \end{center}
    \label{div}
\caption{Snapshots of cell division. The cell
grows by ingesting $X$. Next, the membrane grows inward. 
Finally, the mother cell divides into five daughter cells.}
\end{figure}

\begin{figure}[htbp]
  \begin{center}
    \begin{tabular}{ccc}
      \includegraphics[scale=0.9]{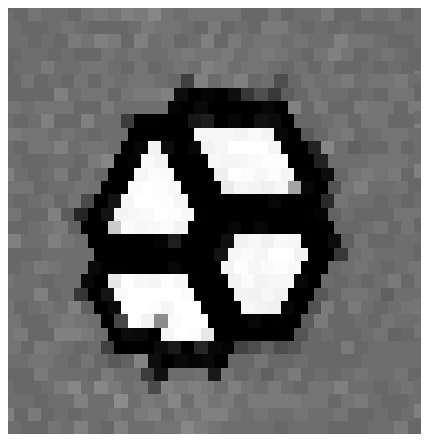} & 
      \includegraphics[scale=0.9]{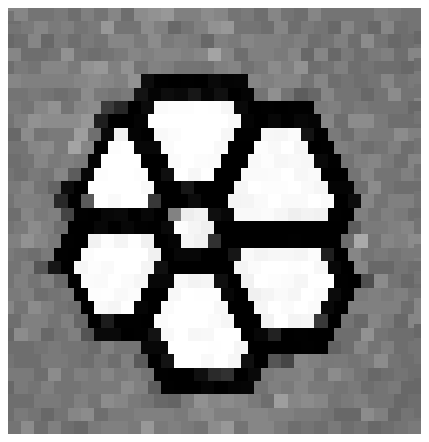} &
      \includegraphics[scale=0.9]{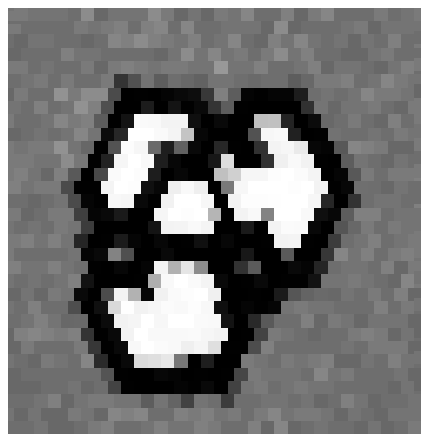} \\
      (a) & (b) & (c)
    \end{tabular}    
  \end{center}
    \label{div2}
\caption{The variations of the cell. The shapes of cells and
their manner of division depend on the values of repulsion
between $M$.
Pictures (a) and (b) show cells with stiff membranes. 
Picture (c) shows cells with flexible membranes. }
\end{figure}


\section{Discussion}

We have demonstrated the model of a self-maintaining cell.
The cell has an internal autocatalytic cycle of chemicals, which
maintains the membrane by itself and the membrane keeps the cell
from collapse. 
We have also shown that the self-maintaining cell can replicate
itself spontaneously; a transition is made from molecular reproduction
to cellular reproduction. 

In real life, the earliest membranes may have been simpler and
rougher than the phospholipid membranes.
The marigranule \cite{Yanagawa1988} represents 
an example of a primitive cell. It has rough shell
and can ingest amino acids from the environment. 
Though the materials of which marigranules consist are very
elementary molecules, there is no linkage between the
organization of shells and its internal dynamics.
If such structure established a symbiotic relationship between 
its embedded chemical network and its membrane, the membranes
could become the targets of the Darwinian selection and evolve into 
more complex structures.

In this study, we have shown that the cell divides in a different
manner according to variation in the 
interaction strength between $M$s. 
It is also true that we can change some properties of biological
membranes 
by varying their components. 
For example,  unsaturated fatty acids components
soften the membrane and cholesterols do the opposite. 

Self-replicating  spots in the
two-dimensional reaction-diffusion system \cite{Pearson1993} 
\cite{Lee1994} have been  well studied.  However, the kinds of
replicating pattern cannot be as diverse as the patterns
generated by cells with membranes. 
The cellular membrane can function as a boundary condition to the
internal chemical network, and conversely, the internal
reactions determine the cell shape. 
In this sense, 
the chemical reactions within cell membranes can be richer than
those without membranes.
The compartmentalization of chemicals will allow cells to be
regarded as units of evolution, because it maintains the
identity of their contents during reproduction.

Koch previously discussed the division mechanisms of
phospholipid vesicles by considering the property of mechanical 
energy \cite{Koch1985}. Our model demonstrates an analogous
dynamic division mechanism. 

The evolution of selective permeability of the membrane must be
considered in future studies. Cell membranes determine how cells
communicate with the environment, including other cells. 
Cells selectively receive stimuli from
the environment and from other 
cells, and respond to these stimuli. 
Our cell model provides several possible approaches by which 
to observe the formation and evolution of membrane functions, and
of how the interaction between cells is generated by each cell's
own internal dynamics. 

\section*{Acknowledgments}
This work is partially supported by
Grant-in aid (No. 09640454) from the Ministry of Education, Science,
Sports and Culture.

\bibliography{nono}

\end{document}